# Exploring the Enablers of Digital Transformation in Small and Medium-Sized Enterprises

**ABSTRACT**

*Recently, digital transformation has caught much attention of both academics and practitioners. With the advent of digital technologies, small-and-medium-sized enterprises (SMEs) have obtained the capacity to initiate digital transformation initiatives in a similar fashion to large-sized organizations. The innate characteristics of digital technologies also favor SMEs in promoting initiation of digital transformation. However, the process digital transformation in SMEs remains a black box and the existing findings of digital transformation in SMEs are limited and remain fragmented. Considering the important contribution SMEs can offer to nations and economies; it is timely and relevant to conduct a profound analysis on digital transformation in SMEs. By conducting a thorough review of existing related literature in management, information systems, and business disciplines, this book chapter aims to understand both internal and external enablers of the digital transformation in SMEs.*

**INTRODUCTION**

In recent times, digital technologies (i.e., social media, mobile technologies, analytics, cloud computing, and internet-of-things), also known as SMAC-IoT have provided myriad prospects for all businesses despite their size (Kraus et al., 2021; Lokuge et al., 2019). For example, in the past, only large-sized organizations with financial capacity had access to resources, the ability to invest in technologies and lead innovation in their respective organizations (Lokuge & Sedera, 2020). As such, technology-led innovation was limited to larger, resourceful organizations (Lokuge & Sedera, 2016; Tan et al., 2017). However, in recent times, digital technologies have disrupted the norms and have given rise to concepts such as digital transformation which has become a buzzword in the business landscape (Vial, 2019).

Digital transformation refers to "the use of digital technologies to create value-added products and processes in organizations and integrate them into business processes, organizational structures, and working models" (Warner and Wäger (2019). As per Vial (2019, p. 118) digital transformation is "a process that aims to improve an entity by triggering significant changes to its properties through combinations of information, computing, communication, and connectivity technologies." Wessel, Baiyere, Ologeanu-Taddei, Cha, and Jensen (2020), extends this conversation and highlights that digital technology plays a key role in digital transformation, whereby they initiate a new organizational identity to develop. Digital transformation enables improved decision-making, value augmentation and improved customer services. The whole objective of investing in digital technologies in organizations is to develop new organizational capabilities for attaining a competitive edge over the market competition (Adikari et al., 2021; Lokuge et al., 2019; Vial, 2019). With the ever-changing customer requirements, advancement of modern technologies and the new normal introduced through COVID-19 pandemic, all organizations, especially small and medium-sized enterprises (SMEs) are interested in digital transformation initiatives considering its benefits.

SMEs are considered as the backbone of the global economy (Deng et al., 2019; Li et al., 2018). For example, in Australia alone 90% of the organizations that contribute to the national income are SMEs (Duan et al., 2012). As such, to increase organizational efficiency, it is important for the SMEs to increase the application of digital technologies and initiate digital transformation (Deng et al., 2019). Therefore, an investigation into the digital transformation in SMEs will assist in uncovering enormous benefits offered through the introduction of digital technologies. As a result of such digital transformations, SMEs will in turn be able to contribute to the economic growth of the community, society, and country.

In defining SMEs, various definitions have been adopted in various countries (Beck et al., 2005; Duan et al., 2012). In general, SMEs are defined as enterprises with less than 250 employees (Beck et al., 2005). SMEs are considered as a distinguished group from large-sized organizations as they possess innate characteristics such as technical incapabilities, limited resources and infrastructure, over influence of the SME owner, lack of control, inadequate capital, high dependence on business partners, informal planning, lack of formal culture, and uncertainty (Deng et al., 2019; King et al., 2014; Nagahawatta et al., 2021). However, on a positive side, SMEs are considered as more agile, flexible, and responsive to market needs (Ghobadian & Gallear, 1997; Li et al., 2018; Sedera & Lokuge, 2017). As such, organizational theories, and day-to-day practices that are relevant for a large-sized organization might not be suitable for a SME (Ghobadian & Gallear, 1997; North et al., 2019; Salim et al., 2015; Szopa & Cyplik, 2020).

Prior research has attempted to unfold the complex process of digital transformation by investigating the strategy of an organization (Bharadwaj et al., 2013; Sedera et al., 2022), its influence on organizational structure (Selander & Jarvenpaa, 2016), business processes (Vial, 2019), and the organizational culture (Karimi & Walter, 2015). Such studies, however, have mostly focused on understanding the digital transformation in large-sized organizations or have taken an "one-size-fits-all" approach regardless of unique features of SMEs. Considering the importance of digital transformation and the unique characteristics of SMEs, it is considered highly timely and relevant to unravel the black box of digital transformation projects among SMEs (Argüelles et al., 2021; Crupi et al., 2020; Gupta & Bose, In Press; Kraus et al., 2021; Lokuge & Duan, 2021; Wessel et al., 2020). In recent times, researchers have seen the value of studying digital transformation in SMEs and have focused on investigating this topic. Even though, there is an interest among scholars in understanding this topic, the extant knowledge regarding the process of digital transformation of SMEs remain limited and disjointed (Ab Wahid & Aziidah Zulkifli, 2021; Garzoni et al., 2020; Li et al., 2018; Lokuge & Duan, 2021). Further, the importance and the unique features of SMEs in the contemporary economic conditions warrant an investigation into the topic of digital transformation in SMEs. As such, a review of the related literature on digital transformation in SMEs can provide a clear understanding of the topic and thereby fill the gap to inform researchers. As such, the proposed overarching research question of this study is derived as follows:

*What are the internal and external enablers of digital transformation in SMEs?*

In order to address this research question, the authors conducted a comprehensive systematic literature review in digital transformation in SMEs. The authors examined 71 articles to identify the enablers of digital transformations in SMEs. The enablers derived through the review of literature provide an early assessment for the SMEs for designing digital strategies for digital transformation.

The structure of this book chapter is as follows. The research method section describes the literature review method followed. The findings of the literature analysis are provided next. Finally, in the conclusion section, future research agenda, limitation and concluding remarks are provided.

## RESEARCH METHOD

A literature review approach was adopted in this study as the objective of this study is to identify internal and external enablers of digital transformation in SMEs. Such an approach is beneficial as (i) it allows the authors to get an insight of the existing literature, and (ii) thereby, it helps authors to determine the research gaps. Five stages were followed for the literature review including defining the scope of the literature review, searching for the related literature, selecting, and finalizing the sample, analyzing the sample, and finalizing the findings (Wolfswinkel et al., 2013).

When defining the scope of the literature review specific inclusion and exclusion criteria were considered. In this study, the authors selected journal papers from the databases such as AIS Library, INFORM, ProQuest, Wiley, Emerald, and EBSCOhost. For a broad coverage of management, business and information systems studies, keywords such as, "digital transformation," OR "digitalization," OR "digitization," OR "digital disruption" AND "Small business," "SME," "Small and medium enterprise" were used for identifying papers. The search was limited to articles published in the English language. The authors only considered articles published before the cut-off date of 16th February 2022 for the sample when the database search was conducted. The analysis was conducted in early February, to ensure all papers considered for the analysis, the authors have considered this cutoff date.

In the next phase, the authors ran the search queries within the chosen databases for obtaining relevant publications for the analysis. Several types of articles were included such as research articles, review articles, opinion pieces, discussion papers, and letters to the editor published in scholarly journals. The search returned 133 articles.

Then, in the next phase, the authors involved screening and selecting the final samples for detailed analysis. The titles and abstracts of all initially identified articles were screened for checking the relevance to the research focus. Duplicate articles were removed. As a result, 80 articles for further review were left. The identified 80 articles have been read in full. Then, the two authors coded the papers. From this, 9 publications were removed as they were not relevant to the topic of analysis. Then, each remaining article (71 articles) identified through the search too was read completely and then determined their relevance. Table 1 below provides details of the sample. Each selected publication was analyzed and coded to identify enablers of the digital transformation process in SMEs.

**Table 1. Overview of the Literature Review Sample**

| Year of publication | Selected papers | Number of papers |
|---|---|---|
| 2022 | (Alam et al. 2022) (Owoseni et al., 2022; Sastararuji et al., 2022) | 3 |
| 2021 | (Ab Wahid & Aziidah Zulkifli, 2021; Akpan et al., 2021; Ghobakhloo & Iranmanesh, 2021; Lindblom et al., 2021; Ratten & Thompson, 2021) | 33 |
| 2020* | (Balakrishnan & Das, 2020; Crupi et al., 2020; Del Giudice et al., 2020; Depaoli et al., 2020; Garzoni et al., 2020; Nwaiwu et al., 2020; Pelletier & Cloutier, 2019; Philipp, 2020; Sanchez-Segura et al., 2020; Szopa & Cyplik, 2020; Wewege et al., 2020; Zangiacomi et al., 2020) | 13 |
| 2019* | (Baber et al., 2019; Bouwman et al., 2019; Chan et al., 2019; Galindo-Martín et al., 2019; Garbellano & Da Veiga, 2019; Llinas & Abad, 2019; Nair et al., 2019; North et al., 2019; Pelletier & Cloutier, 2019; Riera & Iijima, 2019; Sehlin et al., 2019; Ulas, 2019) | 11 |
| 2018 | (Chester Goduscheit & Faullant, 2018; Kim et al., 2018; Lee et al., 2018; Sousa & Wilks, 2018) | 4 |
| 2017 | (Scuotto et al., 2017) | 1 |
| 2016 | (Ansari et al., 2016; Ojala, 2016) | 2 |
| 2015 | (Prindible & Petrick, 2015; Taiminen & Karjaluoto, 2015) | 2 |

| 2014 | (Mehra et al., 2014) | 1 |
| 2013 | (Mathrani et al., 2013) | 1 |

*Only a sample of the references are provided. A full list is available upon request.

## Analysis

The two authors independently analyzed each of the papers identified in the literature sample. When analyzing the sample, authors identified key themes that were emerging. By conducting this approach, the authors were able to maintain an open, and free-flowing mental state that enabled absorbing the phenomenon of interest. All the papers were analyzed separately to code key enablers. When coding the literature sample, the authors labelled any important detail until the existing labels were repeated. Then, each label identified was discussed and explored. The authors refined the coded enablers that were identified independently. Discussions were made until a consensus was made for enablers of digital transformation in SMEs.

## FINDINGS OF THE LITERATURE

This section provides an overview of the analysis of the literature sample. As per the sample, it was evident that the scholars in studying digital transformation in SMEs are focusing more on this topic and this is epitomized by the heightened number of papers published each year from 2013 (only one article) to 2021 (n=35). Among these papers, only a few papers (around 28) focused on studying digital transformation in a particular industry setting such as banking, manufacturing, and IT. When examining the research method adopted in these articles, it was evident that most studies (about 25 papers) followed a qualitative approach, followed by a quantitative approach (n=16), mixed method (n=9), and literature reviews (n=10). Among the papers that conducted a qualitative approach, case studies method was the most applied method and interviews were used to collect qualitative data. Among the sample that applied a quantitative approach, survey method was the most applied approach, followed by econometric modelling. The mixed-method studies mainly applied survey methods along with a few qualitative approaches. In most of the papers in the sample, maturity model as a theoretical framework was applied. In addition, in the sample, commonly used theories and frameworks were absorptive capacity, dynamic capabilities, IS success model, resource-based view, and technology, organization, and environment framework in investigating digital transformation in SMEs.

### Defining Digital Transformation

The accessibility of digital technologies and the recent COVID-19 situation has forced organizations to embrace digital transformation (Argüelles et al., 2021; Vial, 2019). Verhoef et al. (2021, p. 889) define digital transformation as "a change in how a firm employs digital technologies, to develop a new digital business model that helps to create and appropriate more value for the firm." Similarly, researchers such as Fitzgerald et al. (2014) and Lokuge et al. (2021) describe it as the application of new digital technologies to improve the organizational performance in the areas such as developing new business models, customer experience enhancement, and restructuring business operations. Vial (2019, p. 118) define digital transformation as "a process that aims to improve an entity by triggering significant changes to its properties through combinations of information, computing, communication, and connectivity technologies," Wessel et al. (2020), argue that in a digital transformation initiative a new organizational identity is emerged as a result of this transformation. In this process the organizational resources, the practices they follow, and the business strategies they follow radically transformed (Wessel et al., 2020). As per Rowe (2018), digital transformation is "the investment in people and technology to drive a business that is prepared to grow, adapt, scale, and change into the foreseeable future." This shows that digital transformation is a process that employs digital technologies and initiates a positive organizational change in the organization. As per Verhoef et al. (2021), digitization is the starting point of digital transformation process. In digitization, organizations transform analogue data into digitized format. The next step of the digital transformation process, which is digitalization, focuses on automation, value adding and digitalizing business processes. Digital transformation looks at digitalizing business models, ensuring customer and employee experience. In all these endeavors, organizations focus on positive outcomes such as increased organizational performance, increased cost efficiency, competitive advantage, and better customer services (Balakrishnan & Das, 2020; Crupi et al., 2020; Garzoni et al., 2020; Gong et al., 2020; Lokuge, Sedera, Ariyachandra, et al., 2020; Pelletier & Cloutier, 2019; Sedera & Lokuge, 2019; Szopa & Cyplik, 2020).

## Small and Medium-Sized Enterprises

SMEs are considered as an important entity of the global economy. The most important and significant part of the world economy is small businesses (Çakar & Ertürk, 2010; Hung et al., 2004; Lokuge & Duan, 2021). SMEs vary from large organizations regarding their capacity, structure, and size (Tambunan, 2019). The capacity of the SMEs is determined through staff count, assets, size, and revenue. SMEs generally differ from large-sized organizations in terms of their capacity, structure, leadership, finance, and business size (Beck et al., 2005). SMEs possess innate characteristics such as the lack of IT resources, professionals, specialized services or goods, smaller market, and are constantly looking for ways to save costs (Simpson & Docherty, 2004). However, since they have limited staff, communication is efficient, and the ability to execute and decisions making is easy (Simpson & Docherty, 2004).

SMEs are more independent, have less governance structures and are agile (Alsharari et al., 2020; Bouaynaya, 2020; Lokuge & Sedera, 2014b). As per Ghobadian and Gallear (1997), SMEs possess characteristics such as integrated power, multifunctional management, limited product range, flexibility, small management teams, inadequate organizational planning, and unsophisticated IT applications or software. Since they have small teams to manage, SMEs are more risk taking and tend to initiate new products, services and lead digital transformation initiatives. However, due to financial resource limitations, there can be restrictions for successful digital transformation initiatives that they need to manage. This book chapter details both internal and external enablers that SMEs need to consider when initiating digital transformations.

## UNDERSTANDING THE ENABLERS OF DIGITAL TRANSFORMATION

The analysis of literature sample emphasized several important enablers of the digital transformation in SMEs. For example, extant literature highlighted the major role of the changes and the forces in the external environment that forces digital transformation in SMEs. The tension that was arise from continuous changes in the technological landscape, ever increasing market demands, the changes to the lifestyle of the customers and unprecedented events such as the war and the pandemic force digital transformation initiatives to be launched in organizations.

Initially, the authors identified 48 enablers for digital transformation in SMEs from the literature sample. The two authors independently coded and categorized the enablers by merging synonyms and eliminating duplicated codes. In Figure 1, the overview resulted in identifying internal and external enablers of digital transformation in SMEs are shown.

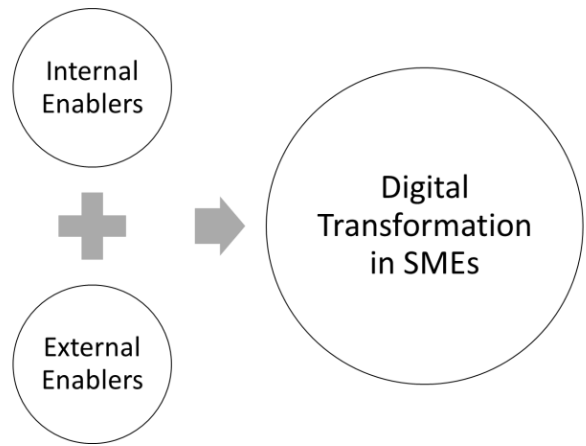

**Figure 1. Key enablers for digital transformation in SMEs**

## Internal Enablers

The analysis of the literature sample derived internal enablers that assisted SMEs in initiating digital transformation initiatives successfully. Figure 2 summarizes the enablers identified through the analysis. The details of each of the enablers are provided below.

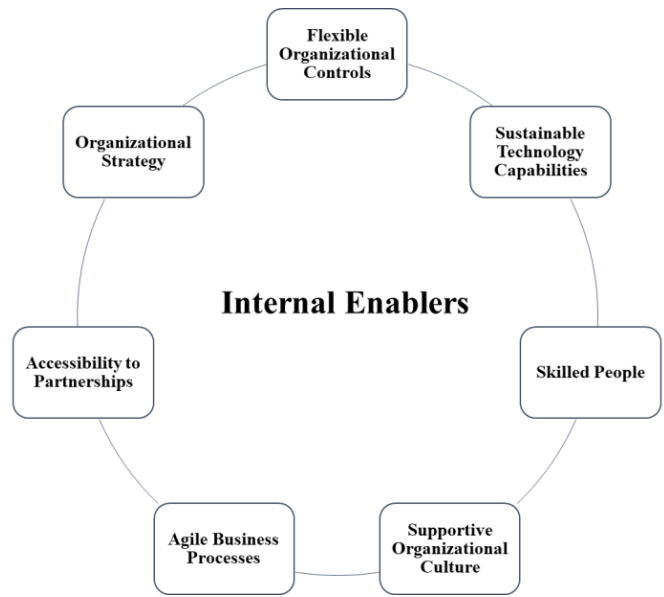

**Figure 2. Internal enablers for digital transformation in SMEs**

**Organizational Strategy:** Organizational strategy is commonly defined as specified goals, objectives, policies, and plans established for operating a particular organization (Miles et al., 1978). Especially for SMEs, most of the articles in the sample emphasized the key role of the clarity of goals and objectives for a SME to conduct digital transformation projects (Nair et al., 2019; Sedera, 2006). For example, having a proper organizational strategy will not affect the digital transformation success. Proper communication of such a strategy is an important element of organizational strategy (Lokuge & Duan, 2021; Lokuge, Sedera, & Palekar, 2020). Considering the size of the SMEs, proper communication of organizational strategy may seem easier. However, it is important to ensure that this communication is properly conducted. When a digital strategy of a particular organization is developed, it is important to identify positioning of their resources, mapping of their capabilities and leveraging such capabilities (Lokuge & Sedera, 2016; Pelletier & Cloutier, 2019). Since SMEs face major resource limitations, it is crucial for the SMEs to determine such capabilities for introducing a strategy that aligns with their requirements (Lokuge & Sedera, 2018, 2020; Lokuge et al., 2016; Nwaiwu et al., 2020).

Further, the criticality of top leadership involvement in managing and ensuring the alignment of such organizational strategies was also considered to be important (Ahmad et al., 2013; Lokuge et al., 2018; Philipp, 2020; Sedera & Dey, 2013; Szopa & Cyplik, 2020). For SMEs, this is one of the most cited enablers that leads to success in digital transformation initiatives. Prior literature agrees that without a proper organizational strategy, strategic initiatives have very little success (Bharadwaj et al., 2013; Henfridsson & Lind, 2014; Nylén & Holmström, 2015; Tan et al., 2015).

**Sustainable Technology Capabilities:** IT capabilities are commonly defined as an organization's ability to construct and use its IT-based resources (including physical IT resources and organization's IT staff), with other organizational resources and capabilities (Bharadwaj et al., 2013; Lokuge & Sedera, 2018) For a SME to ensure positive digital transformation project requires proper utilization of IT capabilities. However, SMEs are inherently possessed limitation in resources (Walther et al., 2013; Walther et al., 2018). Mathrani et al. (2013) highlight that for an SME to be successful in digital transformation projects, it is critical to manage their IT resources and capacities that includes staff, and skills to develop their competencies. Considering the lack of IT resources among SMEs, in instances where they have exclusive access to IT resources, it will determine successful launch of digital transformation initiatives (Philipp, 2020). However, Sedera et al. (2016) highlights that organizations can use their existing IT, such as enterprise systems bundling with complementary IT resources. Especially with the advent of digital technologies, SMEs can afford digital technologies for digital transformation initiatives (Lokuge & Sedera, 2014a). Such endeavors emphasize the criticality of IT knowledge and the capabilities of the staff (Lokuge & Sedera, 2017; Mathrani et al., 2013; Rosemann et al., 2000), which can be challenging for SME. As such, managing technology capabilities determines digital transformation success.

**Accessibility to Partnerships:** As per Ghobakhloo and Iranmanesh (2021) the survival and the success of digital initiatives of the SMEs rely heavily on their capabilities to open up their boundaries to external partners. This includes the partners in their value chain, supply network and new partners introduced through new technologies (Queiroz et al., 2018). The accessibility to leverage the assets, knowledge, and competencies of suppliers, external partners through mergers, alliances, and joint ventures determines the strength of the network (Nuwangi et al., 2018), access to resources and the agility of the SME. In the contemporary digital business landscape, the capacity of the SME is determined through the network of the SME. With the advancement of platform culture, SMEs and all organizations do not necessarily require possessing all resources and need not to have access to resources (Lokuge & Sedera, 2019; Lokuge et al., 2019). If the SMEs have access to a network that possesses the required resources, SMEs can thrive in digital transformation initiatives. In the digital era, with the advent of the shared economy, businesses have the opportunity to engage with new partners. For example, social media provides access to real-time feedback for innovation. In addition, platform providers and mobile app developers create a new ecosystem that provides new opportunities for SMEs. Since SMEs inherently do not have necessary resources and capacity, such alliances and partnerships will create new opportunities to thrive in digital transformation initiatives.

**Flexible Organizational Controls:** SMEs inherently possess no or flexible organizational hierarchy, low degree of formalization, high involvement of the top management, high visibility of the top management, close top management intervention, and high personal authority (Ghobadian & Gallear, 1997). Such characteristics reflect factors that favor innovation (Jansen et al., 2006). As such, in theory, SMEs can easily initiate digital transformation projects as they possess flexibility in organizational structures and controls. Formalization refers to coordinating and controlling the organization (Bodewes, 2002). It is important to maintain a balance between formal and informal controls. Minimal formal coordinating and controlling mechanisms leads to is a SMEs facing negative consequences such as failure of digital transformation projects. As such, while the existing organizational structures and controls of SMEs favor innovation, managing organizational coordination and control are important for SMEs. As such, a balanced organizational structure and control will determine the success of digital transformation initiatives of SMEs.

**Skilled People:** The skilled people enabler entails, knowledge, empowerment, expertise, leadership, and involvement of the managers. Considering digital transformation projects as an innovation to the organization, it can be argued that in all prior literature on innovation the researchers emphasize the importance of managers in leading innovation (Damanpour & Aravind, 2012; Lokuge, Sedera, & Palekar, 2020). As digital transformation process is a radical technological project, managers role is important (Kohli & Melville, 2019). Similarly, in the literature sample most of the papers emphasized the critical role of managers in sharing their knowledge, expertise, leadership, practices that enable successful projects (Crupi et al., 2020). As top management involvement, and support are relatively high in SMEs compared to large organization, digital transformation projects relatively highly successfully in the SME context. In addition, managerial staff are required to educate themselves and reignite their innovation cognition and enhance managerial social capital to promote such projects (Li et al., 2018). As a result, the manager's knowledge and capacity determine the success of digital transformation projects. Further, possessing skilled people is also associated with accessibility to partnerships as partnerships enable existence of skilled people.

**Agile Business Processes:** Organizational business processes being agile is important for the success of digital transformation projects. In particular, the agility of such business processes determines the success of digital transformation initiatives. As per Markus and Tanis (2000), implementation of an enterprise system was a radical innovation that re-engineers the organizational business processes. Similarly, digital transformation processes can restructure business processes and develop a new organizational identity (Depaoli et al., 2020). Even though such radical changes are important, it is mandatory to manage such radical changes carefully. Especially for SMEs, managing the business processes carefully becomes one of the prime factors as most SMEs do not possess standard, consistent business processes existing in their organizations. Such irregularities may hinder the digital transformation projects in SMEs. However, agility in redeveloping the organizational business processes is key for success of digital transformation endeavors. This is evident in the COVID-19 pandemic where organization agility emerges as a key enabler in driving digital transformation in SMEs to counter the uncertain environment (Troise et al. 2022).

**Organizational Culture:** As per Wood (2001), organizational culture is defined as the systems of shared beliefs and values that develop within an organization or within its sub-units and that guides the behavior of its members. As per extant literature organizational culture is considered as one of the most important factors that establish the extent of innovation in organizations (Adams et al., 2006; Boudreau & Lakhani, 2013; Büschgens et al., 2013; Lee et al., 2016). This is especially true for SMEs as considering its size and the limited resources availability, it is very important for the managers to ensure positive innovation favoring culture in the organization. Such innovation-favoring environment will ensure the success of digital transformation projects as well. When considering SME context, commonly studied cultural dimensions like empowerment, collectivism, uncertainty avoidance, power distance highlight positives of SMEs that enable digital transformation in the organization (Çakar & Ertürk, 2010). Organizational culture guides the organizational business processes as well as controls. As such, proper management and establishing enriching organizational culture will determine the success of digital transformation initiatives in SMEs.

## External Enablers

Three external enablers (as displayed in Figure 3) were identified from the literature analysis that have key influence on driving the digital transformation in SMEs. While the authors acknowledge additional enablers such as governmental interventions, policies and rules as factors that may influence digital transformation in organizations, the findings are limited as the book chapter reports only the findings of the literature analysis sample. The details of each of the enablers are provided below.

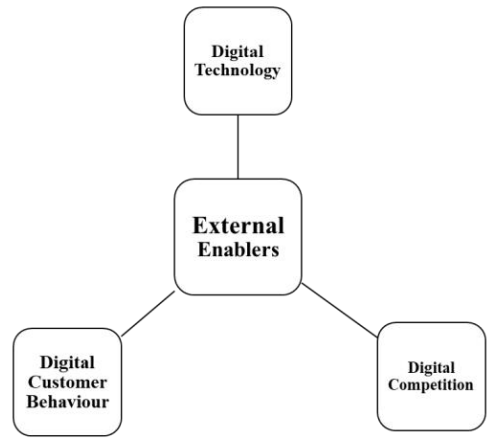

**Figure 3. External enablers for digital transformation in SMEs**

**Digital Technology:** Nambisan (2013) identified digital technologies as new IT-based resources that include social media, mobile technologies, analytics, cloud computing, and internet-of-things. In the contemporary technology landscape, in addition to new changes to these existing technologies, new and advanced technologies are on the rise. For example, technologies such as robotics, nanotechnologies, blockchain and cognitive technologies are changing the very nature of how organizations conduct their businesses. Such changes are demanding organizations to be innovative and create a competitive business landscape. As such, digital technology becomes an external enabler that drives digital transformation initiatives. However, internally, SMEs need to consider the expertise and financial capacity to manage these technologies before investing in digital technologies for digital transformation initiatives. They also need to understand how to adapt and integrate digital technologies with business functions, so that they can leverage their competitiveness through digital transformation. While the innate characteristics of digital technologies such as cost efficiency, ease-of-use, ease-of-implementation favors SMEs, it is important to understand the market and establish a strategy that aligns with the organizational capacity.

**Digital Competition:** The advancement of social media and mobile technologies have changed the way organizations connect with their customers. The advent of these technologies has provided even micro-sized organizations with an opportunity to compete with the rest of the organizations. As such, in this modern day and age, there is extremely fierce competition among organizations. The price of the goods and services, the availability, delivery times of the goods and services no longer provide a competitive advantage for the organizations. As per the Darwinian law, the 'digitally strongest' organizations thrive and survive in this digital competition. With the capacities provided through social media, SMEs can understand market needs. The real-time feedback offered via such platforms enables SMEs to make changes to their products and services. Not only for obtaining feedback, but SMEs are also able to fiercely market their products and services via these channels. Previously, organizations were required to allocate an extensive budget for marketing and promotion. However, such initiatives have been competitive with the advent of digital technologies. Therefore, this is considered as an external driver for digital transformation initiatives in SMEs.

**Digital Customer Behavior:** The contemporary business landscape has changed immensely due to the advent of digital technologies. The COVID-19 pandemic has exacerbated this situation and as a result, how organizations operate has changed as well. For example, before the pandemic, remote working was not a prominent concept. In addition, e-commerce was common, however, among SMEs this was not a popular concept. For the survival of the SMEs, it was mandatory for them to have a digital presence and build customer agility (Troise et al. 2022). Customer agility is the extent to which a SME senses and responds to customer needs (Atapattu & Sedera, 2014). SMEs are in a better position to obtain knowledge regarding market opportunities when they utilize the feedback of customers and involve them in creating new ideas, products, and services. Such digital transformation initiatives were enforced due to digital customer behaviors. The advent of concepts such as doing business via social media sites, social media influencers have transformed the very nature of modern businesses. The lifestyles, routines, norms of the customers are invariably intertwined with digital technologies. As such, customer behaviors too, have changed due to these changes. As such, digital customer behavior is considered as an external driver of digital transformation among SMEs.

## CONCLUSION AND FUTURE WORK

The objective of this book chapter was to understand the enablers of digital transformation in SMEs. To analyze this phenomenon, in this book chapter, the authors conducted a systematic review of 71 papers in multiple disciplines that discussed digital transformation in SMEs. The literature analysis allowed the revelation of internal and external enablers for digital transformation in SMEs. Specifically, internal factors such as facilitating organizational strategy, sustainable technology capabilities, flexible organizational controls, accessibility to partnerships, agile business processes, skilled people, and supportive organizational culture facilitate digital transformation in SMEs. External factors including available digital technologies, digital customer behavior, and digital competition are important considerations for SMEs in their digital transformation journey.

This research provides both theoretical and practical contributions to the field. Theoretically, this book chapter opens new pathways for researchers to better understand digital transformation in SMEs. While much research has been done to unpack the black box of digital transformation process in large organizations (Bharadwaj et al., 2013; Selander & Jarvenpaa, 2016; Vial, 2019), limited insights have been provided on understanding digital transformation process in SMEs. The advent of digital technologies has offered SMEs the opportunity to initiate digital transformation in a similar fashion as large-sized organizations. SMEs with unique characteristics, however, warrant the need for a separate investigation. This study hence contributes by providing insights into digital transformation in SMEs through the identification of key enablers. Next, this study provides two frameworks to summarize these key enablers for facilitating digital transformation in SMEs. Such frameworks serve as the basis for future research to be tested and validated to provide empirical evidence. Practically, the findings of this study will be important for policy makers, and managers of SMEs to understand and determine the factors for facilitating successful digital transformation in SMEs. SMEs have been playing a key role in the national economy, a profound understanding of the digital transformation process in SMEs and its enablers will provide enormous benefits for SMEs to bring in digital technologies for their organizations.

There are certain limitations for this study. Since this research is at the conceptual level, empirical research is required to further establish the findings. For example, while there can be more new enablers of digital transformation in SMEs, the findings are limited to the existing studies. Future research can focus on validating the suggested enablers and thereby assisting SMEs to design effective digital strategies. Next, the authors acknowledge the size of the literature sample as the authors only considered journals for our analysis. Recent quality publications such as conference proceedings can be included in future research to extend the scope of the review and enrich the findings. In addition, there can be associations among the enablers. Such nuance associations can only be identified through an explorative study. The authors encourage future studies to be conducted on this line to contribute to both academia and practice.

## REFERENCES

Ab Wahid, R., & Aziidah Zulkifli, N. (2021). Factors Affecting the Adoption of Digital Transformation among SME's in Malaysia. *Journal of Information Technology Management*, *13*(3), 126-140.

Adams, R., Bessant, J., & Phelps, R. (2006). Innovation management measurement: A review. *International Journal of Management Reviews*, *8*(1), 21-47. https://doi.org/10.1111/j.1468-2370.2006.00119.x

Adikari, A., Burnett, D., Sedera, D., de Silva, D., & Alahakoon, D. (2021). Value co-creation for open innovation: An evidence-based study of the data driven paradigm of social media using machine learning. *International Journal of Information Management Data Insights*, *1*(2), 100022.

Ahmad, M. N., Zakaria, N. H., & Sedera, D. (2013). Ontology-based knowledge management for enterprise systems. In *Competition, Strategy, and Modern Enterprise Information Systems* (pp. 184-212). IGI Global.

Akpan, I. J., Soopramanien, D., & Kwak, D.-H. (2021). Cutting-edge technologies for small business and innovation in the era of COVID-19 global health pandemic. *Journal of Small Business & Entrepreneurship*, *33*(6), 607-617.

Alsharari, N. M., Al-Shboul, M., & Alteneiji, S. (2020). Implementation of cloud ERP in the SME: evidence from UAE. *Journal of Small Business and Enterprise Development*, *27*(2), 299-327. https://doi.org/10.1108/JSBED-01-2019-0007

Ansari, S., Garud, R., & Kumaraswamy, A. (2016). The disruptor's dilemma: TiVo and the US television ecosystem. *Strategic Management Journal*, *37*(9), 1829-1853.

Argüelles, A. J., Cortés, H. D., Ramirez, O. E. P., & Bustamante, O. A. (2021). Technological Spotlights of Digital Transformation: Uses and Implications Under COVID-19 Conditions. In *Information Technology Trends for a Global and Interdisciplinary Research Community* (pp. 19-49). IGI Global.

Atapattu, M., & Sedera, D. (2014). Agility in consumer retail: Sense-Response Alignment through the eyes of customers. *Australasian Journal of Information Systems*, *18*(2), 111-132.

Baber, W. W., Ojala, A., & Martinez, R. (2019). Effectuation logic in digital business model transformation: Insights from Japanese high-tech innovators. *Journal of Small Business and Enterprise Development*, *26*(6/7), 811-830.

Balakrishnan, R., & Das, S. (2020). How do firms reorganize to implement digital transformation? *Strategic Change*, *29*(5), 531-541.

Beck, R., Wigand, R. T., & König, W. (2005). The diffusion and efficient use of electronic commerce among small and medium-sized enterprises: an international three-industry survey. *Electronic Markets*, *15*(1), 38-52.

Bharadwaj, A., El Sawy, O. A., Pavlou, P. A., & Venkatraman, N. (2013). Digital business strategy: toward a next generation of insights. *MIS quarterly*, *37*(2), 471-482.

Bodewes, W. E. J. (2002). Formalization and innovation revisited. *European Journal of Innovation Management*, *5*(4), 214-223. https://doi.org/doi:10.1108/14601060210451171

Bouaynaya, W. (2020). Cloud computing in SMEs: towards delegation of the CIO role. *Information & Computer Security*, *28*(2), 199-213. https://doi.org/10.1108/ICS-01-2017-0001

Boudreau, K. J., & Lakhani, K. R. (2013). Using the crowd as an innovation partner. *Harvard business review*, *91*(4), 60-69.

Bouwman, H., Nikou, S., & de Reuver, M. (2019). Digitalization, business models, and SMEs: How do business model innovation practices improve performance of digitalizing SMEs? *Telecommunications Policy*, *43*(9), 101828.

Büschgens, T., Bausch, A., & Balkin, D. B. (2013). Organizational Culture and Innovation: A Meta-Analytic Review. *Journal of product innovation management*, *30*(4), 763-781.

Çakar, N. D., & Ertürk, A. (2010). Comparing innovation capability of small and medium-sized enterprises: examining the effects of organizational culture and empowerment. *Journal of Small Business Management*, *48*(3), 325-359.

Chan, C. M., Teoh, S. Y., Yeow, A., & Pan, G. (2019). Agility in responding to disruptive digital innovation: Case study of an SME. *Information Systems Journal*, *29*(2), 436-455.

Chester Goduscheit, R., & Faullant, R. (2018). Paths toward radical service innovation in manufacturing companies—A service-dominant logic perspective. *Journal of product innovation management*, *35*(5), 701-719.

Crupi, A., Del Sarto, N., Di Minin, A., Gregori, G. L., Lepore, D., Marinelli, L., & Spigarelli, F. (2020). The digital transformation of SMEs–a new knowledge broker called the digital innovation hub. *Journal of Knowledge Management*, *24*(6), 1263-1288.

Damanpour, F., & Aravind, D. (2012). Managerial Innovation: Conceptions, Processes, and Antecedents. *Management and Organization Review*, *8*(2), 423-454.

Del Giudice, M., Scuotto, V., Papa, A., Tarba, S., Bresciani, S., & Warkentin, M. (2020). A self-tuning model for smart manufacturing SMEs: Effects on digital innovation. *Journal of product innovation management*, *38*(1), 68-89.

Deng, H., Duan, S. X., & Luo, F. (2019). Critical determinants for electronic market adoption: Evidence from Australian small-and medium-sized enterprises. *Journal of Enterprise Information Management*, *33*(2), 335-352.

Depaoli, P., Za, S., & Scornavacca, E. (2020). A model for digital development of an interaction-based approach. *Journal of Small Business and Enterprise Development*, *27*(7), 1049-1068.

Fitzgerald, M., Kruschwitz, N., Bonnet, D., & Welch, M. (2014). Embracing digital technology: A new strategic imperative. *MIT Sloan Management Review*, *55*(2), 1-16.

Galindo-Martín, M.-Á., Castaño-Martínez, M.-S., & Méndez-Picazo, M.-T. (2019). Digital transformation, digital dividends and entrepreneurship: A quantitative analysis. *Journal of business research*, *101*, 522-527.

Garbellano, S., & Da Veiga, M. d. R. (2019). Dynamic capabilities in Italian leading SMEs adopting industry 4.0. *Measuring Business Excellence*, *23*(4), 472-483.

Garzoni, A., De Turi, I., Secundo, G., & Del Vecchio, P. (2020). Fostering digital transformation of SMEs: a four levels approach. *Management Decision*, *58*(8), 1543-1562.

Ghobadian, A., & Gallear, D. (1997). TQM and organization size. *International journal of operations & production management*, *17*(2), 121-163.

Ghobakhloo, M., & Iranmanesh, M. (2021). Digital transformation success under Industry 4.0: A strategic guideline for manufacturing SMEs. *Journal of Manufacturing Technology Management*, *32*(8), 1533-1556.

Gong, Y., Yang, J., & Shi, X. (2020). Towards a comprehensive understanding of digital transformation in government: Analysis of flexibility and enterprise architecture. *Government Information Quarterly*, *37*(3), 1-13.

Gupta, G., & Bose, I. (In Press). Digital transformation in entrepreneurial firms through information exchange with operating environment. *Information & Management*.

Henfridsson, O., & Lind, M. (2014). Information systems strategizing, organizational sub-communities, and the emergence of a sustainability strategy. *The Journal of Strategic Information Systems*, *23*(1), 11-28.

Hung, S.-Y., Chang, S.-I., & Lee, P.-J. (2004, July 08-11). Critical Factors Of ERP Adoption For Small- And Medium- Sized Enterprises: An Empirical Study. Proceedings of the 8th Pacific Asia Conference on Information Systems, Shanghai, China.

Jansen, J. J. P., Van Den Bosch, F. A. J., & Volberda, H. W. (2006). Exploratory Innovation, Exploitative Innovation, and Performance: Effects of Organizational Antecedents and Environmental Moderators. *Management Science*, *52*(11), 1661-1674.

Karimi, J., & Walter, Z. (2015). The Role of Dynamic Capabilities in Responding to Digital Disruption: A Factor-Based Study of the Newspaper Industry. *Journal of management information systems*, *32*(1), 39-81. https://doi.org/10.1080/07421222.2015.1029380

Kim, Y. J., Kim, K.-C., Song, C. S., & Kang, M. S. (2018). The impact of humane entrepreneurship on business ecosystem and economic development. *The Journal of Small Business Innovation*, *20*(4), 3-10.

King, B. E., Breen, J., & Whitelaw, P. A. (2014). Hungry for growth? Small and medium-sized tourism enterprise (SMTE) business ambitions, knowledge acquisition and industry engagement. *International Journal of Tourism Research*, *16*(3), 272-281.

Kohli, R., & Melville, N. P. (2019). Digital innovation: A review and synthesis. *Information Systems Journal*, *29*(1), 200-223.

Kraus, S., Jones, P., Kailer, N., Weinmann, A., Chaparro-Banegas, N., & Roig-Tierno, N. (2021). Digital transformation: An overview of the current state of the art of research. *SAGE Open*, *11*(3), 1-15.

Lee, C.-S., Kim, Y.-K., & Kim, S.-H. (2018). A study on the support policy for digital transformation of small businesses. *Journal of Distribution Science*, *16*(2), 89-99.

Lee, M. T., Raschke, R. L., & Louis, R. S. (2016). Exploiting organizational culture: Configurations for value through knowledge worker's motivation. *Journal of business research*, *69*(11), 5442–5447.

Li, L., Su, F., Zhang, W., & Mao, J. Y. (2018). Digital transformation by SME entrepreneurs: A capability perspective. *Information Systems Journal*, *28*(6), 1129-1157.

Lindblom, P., Nygren, E., Kolog, E. A., & Sutinen, E. (2021). Defining 'Smart Rural'in the Framework of Regional Digitalisation. 2021 IST-Africa Conference (IST-Africa),

Llinas, D., & Abad, J. (2019). The role of high-performance people management practices in Industry 4.0: The case of medium-sized Spanish firms. *Intangible Capital*, *15*(3), 190-207.

Lokuge, S., & Duan, S. X. (2021). *Towards Understanding Enablers of Digital Transformation in Small and Medium-Sized Enterprises* Australasian Conference on Information Systems,

Lokuge, S., & Sedera, D. (2014a). Deriving Information Systems Innovation Execution Mechanisms. Australasian Conference on Information Systems, Auckland, New Zealand.

Lokuge, S., & Sedera, D. (2014b). Enterprise systems lifecycle-wide innovation readiness. Pacific Asia Conference on Information Systems, Chengdu, China.

Lokuge, S., & Sedera, D. (2016). Is your IT eco-system ready to facilitate organizational innovation? Deriving an IT eco-system readiness measurement model. International Conference on Information Systems, Dublin, Ireland.

Lokuge, S., & Sedera, D. (2017). Turning Dust to Gold: How to increase inimitability of Enterprise System. Pacific Asia Conference on Information Systems, Langkawi, Malaysia.

Lokuge, S., & Sedera, D. (2018). The Role of Enterprise Systems in Fostering Innovation in Contemporary Firms. *Journal of Information Technology Theory and Application (JITTA)*, *19*(2), 7-30.

Lokuge, S., & Sedera, D. (2019). *Attaining business alignment in information technology innovations led by line-of-business managers* Australasian Conference on Information Systems, Perth, Australia.

Lokuge, S., & Sedera, D. (2020). Fifty Shades of Digital Innovation: How Firms Innovate with Digital Technologies. Pacific Asia Conference on Information Systems, Dubai, UAE.

Lokuge, S., Sedera, D., Ariyachandra, T., Kumar, S., & Ravi, V. (2020). The Next Wave of CRM Innovation: Implications for Research, Teaching, and Practice. *Communications of the Association for Information Systems*, *46*(1), 560-583.

Lokuge, S., Sedera, D., Cooper, V., & Burstein, F. (2021). Digital Transformation: Environmental Friend or Foe? Panel Discussion at the Australasian Conference on Information Systems 2019. *Communications of the Association for Information Systems*, *48*, 47.

Lokuge, S., Sedera, D., & Grover, V. (2016). Thinking inside the box: Five Organizational strategies enabled through Information Systems. Pacific Asia Conference on Information Systems, Chiyai, Taiwan.

Lokuge, S., Sedera, D., Grover, V., & Xu, D. (2019). Organizational readiness for digital innovation: Development and empirical calibration of a construct. *Information & Management*, *56*(3), 445-461.

Lokuge, S., Sedera, D., & Palekar, S. (2020). The Clash of the Titans: CIO and LOB Engagement in IT Innovation. In K. Sandhu (Ed.), *Leadership, Management, and Adoption Techniques for Digital Service Innovation* (pp. 86-102). IGI Global.

Lokuge, S., Sedera, D., & Perera, M. (2018). The Clash of the Leaders: The intermix of leadership styles for resource bundling. Pacific Asia Conference on Information Systems, Yokohama, Japan.

Markus, L., & Tanis, C. (2000). The Enterprise Systems Experience - from Adoption to Success. In R. W. Zmud (Ed.), *Framing the Domains of IT Management: Projecting the Future through the Past* (pp. 173-207). Pinnaflex Educational Resources, Inc. http://www.pinnaflex.com/pdf/framing/CH10.pdf

Mathrani, S., Mathrani, A., & Viehland, D. (2013). Using enterprise systems to realize digital business strategies. *Journal of Enterprise Information Management*, *26*(4), 363-386.

Mehra, A., Langer, N., Bapna, R., & Gopal, R. (2014). Estimating Returns to Training in the Knowledge Economy. *MIS quarterly*, *38*(3), 757-772.

Miles, R. E., Snow, C. C., Meyer, A. D., & Coleman, H. J. (1978). Organizational strategy, structure, and process. *Academy of management review*, *3*(3), 546-562.

Nagahawatta, R., Warren, M., Lokuge, S., & Salzman, S. (2021). Security concerns influencing the adoption of cloud computing of SMEs: a literature review. Americas Conference on Information Systems,

Nair, J., Chellasamy, A., & Singh, B. B. (2019). Readiness factors for information technology adoption in SMEs: testing an exploratory model in an Indian context. *Journal of Asia Business Studies*, *13*(4), 694-718.

Nambisan, S. (2013). Information Technology and Product/Service Innovation: A Brief Assessment and Some Suggestions for Future Research. *Journal of the Association for Information Systems*, *14*(4), 215-226.

North, K., Aramburu, N., & Lorenzo, O. J. (2019). Promoting digitally enabled growth in SMEs: a framework proposal. *Journal of Enterprise Information Management*, *33*(1), 238-262.

Nuwangi, S. M., Sedera, D., & Srivastava, S. C. (2018). *Multi-layered control mechanisms in software development outsourcing* Pacific Asia Conference on Information Systems, Japan

Nwaiwu, F., Duduci, M., Chromjakova, F., & Otekhile, C.-A. F. (2020). Industry 4.0 concepts within the czech sme manufacturing sector: An empirical assessment of critical success factors. *Business: Theory and Practice*, *21*(1), 58-70.

Nylén, D., & Holmström, J. (2015). Digital innovation strategy: A framework for diagnosing and improving digital product and service innovation. *Business Horizons*, *58*(1), 57-67.

Ojala, A. (2016). Business models and opportunity creation: How IT entrepreneurs create and develop business models under uncertainty. *Information Systems Journal*, *26*, 451–476.

Owoseni, A., Hatsu, S., & Tolani, A. (2022). How do digital technologies influence the dynamic capabilities of micro and small businesses in a pandemic and low-income country context? *The Electronic Journal of Information Systems in Developing Countries*, *88*(2), e12202.

Pelletier, C., & Cloutier, L. M. (2019). Conceptualising digital transformation in SMEs: an ecosystemic perspective. *Journal of Small Business and Enterprise Development*, *26*(6/7), 855-876.

Philipp, R. (2020). Digital readiness index assessment towards smart port development. *NachhaltigkeitsManagementForum*, *28*, 49-60.

Prindible, M., & Petrick, I. (2015). Learning the building blocks of service innovation from SMEs. *Research-Technology Management*, *58*(5), 61-63.

Ratten, V., & Thompson, A.-J. (2021). Digital transformation from COVID-19 in small business and sport entities. In *COVID-19 and entrepreneurship* (pp. 54-70). Routledge.

Riera, C., & Iijima, J. (2019). The role of IT and organizational capabilities on digital business value. *Pacific Asia Journal of the Association for Information Systems*, *11*(2), 67-95.

Rosemann, M., Sedera, W., & Sedera, D. (2000). Industry-oriented education in enterprise systems. Australasian Conference on Information Systems, Brisbane, Australia.

Rowe, F. (2018). Being critical is good, but better with philosophy! From digital transformation and values to the future of IS research. *European Journal of Information Systems*, *27*(3), 380-393.

Salim, S. A., Sedera, D., Sawang, S., Alarifi, A. H. E., & Atapattu, M. (2015). Moving from Evaluation to Trial: How do SMEs Start Adopting Cloud ERP? *Australasian Journal of Information Systems*, *19*, S219-S254.

Sanchez-Segura, M. I., Medina-Dominguez, F., de Amescua, A., & Dugarte-Peña, G. L. (2020). Knowledge governance maturity assessment can help software engineers during the design of business digitalization projects. *Journal of Software: Evolution and Process*, 1-25.

Sastararuji, D., Hoonsopon, D., Pitchayadol, P., & Chiwamit, P. (2022). Cloud accounting adoption in Thai SMEs amid the COVID-19 pandemic: an explanatory case study. *Journal of Innovation and Entrepreneurship*, *11*(1), 1-25.

Scuotto, V., Caputo, F., Villasalero, M., & Del Giudice, M. (2017). A multiple buyer–supplier relationship in the context of SMEs' digital supply chain management. *Production Planning & Control*, *28*(16), 1378-1388.

Sedera, D. (2006). An empirical investigation of the salient characteristics of IS-Success models. Americas Conference on Information Systems, Acapulco, Mexico.

Sedera, D., & Dey, S. (2013). User expertise in contemporary information systems: Conceptualization, measurement and application. *Information & Management 50*(8), 621–637

Sedera, D., & Lokuge, S. (2017). The Role of Enterprise Systems in Innovation in the Contemporary Organization. In R. G. Galliers & M.-K. Stein (Eds.), *The Routledge Companion to Management Information Systems* (pp. 608). The Routledge

Sedera, D., & Lokuge, S. (2019). *Do we put all eggs in one basket? A polynomial regression study of digital technology configuration strategies* International Conference on Information Systems, Munich, Germany.

Sedera, D., Tan, C.-W., & Xu, D. (2022). Digital business transformation in innovation and entrepreneurship. *Information & Management*, *59*(3), 103620.

Sehlin, D., Truedsson, M., & Cronemyr, P. (2019). A conceptual cooperative model designed for processes, digitalisation and innovation. *International Journal of Quality and Service Sciences*, *11*(4), 504-522.

Selander, L., & Jarvenpaa, S. L. (2016). Digital Action Repertoires and Transforming a Social Movement Organization. *MIS quarterly*, *40*(2), 331-352.

Simpson, M., & Docherty, A. J. (2004). E-commerce adoption support and advice for UK SMEs. *Journal of Small Business and Enterprise Development*, *11*(3), 315-328.

Sousa, M. J., & Wilks, D. (2018). Sustainable skills for the world of work in the digital age. *Systems Research and Behavioral Science*, *35*(4), 399-405.

Szopa, Ł., & Cyplik, P. (2020). The concept of building a digital transformation model for enterprises from the SME sector. *LogForum*, *16*(4), 593-601.

Taiminen, H. M., & Karjaluoto, H. (2015). The usage of digital marketing channels in SMEs. *Journal of Small Business and Enterprise Development*, *22*(4), 633-651.

Tambunan, T. (2019). Recent evidence of the development of micro, small and medium enterprises in Indonesia. *Journal of Global Entrepreneurship Research*, *9*(1), 1-15.

Tan, B., Pan, S. L., Lu, X., & Huang, L. (2015). The Role of IS Capabilities in the Development of Multi-Sided Platforms: The Digital Ecosystem Strategy of Alibaba. com. *Journal of the Association for Information Systems*, *16*(4), 248-280.

Tan, F. T. C., Tan, B., Wang, W., & Sedera, D. (2017). Management Innovation for IT-Enabled Operational Agility: An Interdependencies Perspective. *Information & Management*, *54*(3), 292-303.

Ulas, D. (2019). Digital transformation process and SMEs. *Procedia computer science*, *158*, 662-671.

Verhoef, P. C., Broekhuizen, T., Bart, Y., Bhattacharya, A., Dong, J. Q., Fabian, N., & Haenlein, M. (2021). Digital transformation: A multidisciplinary reflection and research agenda. *Journal of business research*, *122*, 889-901.

Vial, G. (2019). Understanding digital transformation: A review and a research agenda. *The Journal of Strategic Information Systems*, *28*(2), 118-144.

Walther, S., Sarker, S., Sedera, D., & Eymann, T. (2013). Exploring Subscription Renewal Intention Of Operational Cloud Enterprise Systems-A Socio-Technical Approach. European Conference on Information Systems, Utrecht, The Netherlands.

Walther, S., Sedera, D., Urbach, N., Eymann, T., Otto, B., & Sarker, S. (2018). Should We Stay, or Should We Go? Analyzing Continuance of Cloud Enterprise Systems. *Journal of Information Technology Theory and Application (JITTA)*, *19*(2), 57-88.

Warner, K. S., & Wäger, M. (2019). Building dynamic capabilities for digital transformation: An ongoing process of strategic renewal. *Long Range Planning*, *52*(3), 326-349.

Wessel, L., Baiyere, A., Ologeanu-Taddei, R., Cha, J., & Jensen, T. (2020). Unpacking the difference between digital transformation and IT-enabled organizational transformation. *Journal of Association of Information Systems*.

Wewege, L., Lee, J., & Thomsett, M. C. (2020). Disruptions and Digital Banking Trends. *Journal of Applied Finance and Banking*, *10*(6), 15-56.

Wolfswinkel, J. F., Furtmueller, E., & Wilderom, C. P. (2013). Using grounded theory as a method for rigorously reviewing literature. *European Journal of Information Systems*, *22*(1), 45-55.

Zangiacomi, A., Pessot, E., Fornasiero, R., Bertetti, M., & Sacco, M. (2020). Moving towards digitalization: a multiple case study in manufacturing. *Production Planning & Control*, *31*(2-3), 143-157.

**ADDITIONAL READING**

Gaddefors, J., and Anderson, A.R. 2019. "Romancing the Rural: Reconceptualizing Rural Entrepreneurship as Engagement with Context (S)," *The International Journal of Entrepreneurship and Innovation* (20:3), pp 159-169.

Gupta, A.K. 2013. "Tapping the Entrepreneurial Potential of Grassroots Innovation," *Stanford Social Innovation Review* (11:3), pp 18-20.

Kohli, R., and Melville, N.P. 2019. "Digital Innovation: A Review and Synthesis," *Information Systems Journal* (29:1), pp 200-223.

Nambisan, S. 2017. "Digital Entrepreneurship: Toward a Digital Technology Perspective of Entrepreneurship," *Entrepreneurship Theory and Practice* (41:6), pp 1029-1055.

Nambisan, S., Wrig*ht, M., and Feldman, M. 2019.* "The Digital Transformation of Innovation and Entrepreneurship: Progress, Challenges and Key Themes," Research Policy (48:8), pp 1-9.

Vial, G. 2019. "Understanding Digital Transformation: A Review and a Research Agenda," The Journal of Strategic Information Systems (28:2), pp 118-144.

Wessel, L., Baiyere, A., Ologeanu-Taddei, R., Cha, J., and Jensen, T. 2020. "Unpacking the Difference between Digital Transformation and It-Enabled Organizational Transformation," *Journal of Association of Information Systems*).

Yoo, Y., Boland Jr, R.J., Lyytinen, K., and Majchrzak, A. 2012. "Organizing for Innovation in the Digitized World," *Organization Science* (23:5), pp 1398-1408.

**KEY TERMS AND DEFINITIONS**

**Digital technologies:** new technologies such as social media, mobile technologies, analytics, cloud computing and the internet-of-things are considered as digital technologies.

**Small and Medium sized Enterprises:** enterprises with less than 250 employees

**Digital Transformation:** the adoption of digital technology by an organization to improve organizational efficiency, attain competitive advantage.

**Literature review:** a system analysis of published papers intended to explain the extant understanding of a topic.

**Framework:** a conceptual structure for explaining a phenomenon

**Entrepreneurship:** the discovery, evaluation, and exploitation of future goods and services

**Innovation:** implementation of an idea whether pertaining to a device, system, process, policy, program, or service that is new to the organization.